# Fast MRI of bones in the knee – An AI-driven reconstruction approach for adiabatic inversion recovery prepared ultra-short echo time sequences

**Authors**

Philipp Hans Nunn[a,*], Henner Huflage[a], Jan-Peter Grunz[a], Philipp Gruschwitz[a], Oliver Schad[a], Thorsten Alexander Bley[a], Johannes Tran-Gia[b], Tobias Wech[a,c]

**\*Corresponding author at:**

Department of Diagnostic and Interventional Radiology
University Hospital Würzburg
Oberdürrbacher Str. 6, Würzburg, 97080, Germany
E-Mail: Nunn_P1@ukw.de

[a] Department of Diagnostic and Interventional Radiology, University Hospital Würzburg, Oberdürrbacher Straße 6, Würzburg, 97080, Germany
[b] Department of Nuclear Medicine, University Hospital Würzburg, Oberdürrbacher Straße 6, Würzburg, 97080, Germany
[c] Comprehensive Heart Failure Center Würzburg, University Hospital Würzburg, Am Schwarzenberg 15, 97078 Würzburg, Germany

## Acknowledgements

This work was supported by the Interdisciplinary Center for Clinical Research in Würzburg (Grant F-437), the German Research Foundation (DFG, 521044818), the German Federal Ministry of Education and Research (BMBF, 13GW0357B), and the project 22HLT03 AlphaMet, which has received funding from the European Partnership on Metrology, co-financed from the European Union's Horizon Europe Research and Innovation Programme and by the Participating States.

# Abstract

**Purpose:** Inversion recovery prepared ultra-short echo time (IR-UTE)-based MRI enables radiation-free visualization of osseous tissue. However, sufficient signal-to-noise ratio (SNR) can only be obtained with long acquisition times. This study proposes a data-driven approach to reconstruct undersampled IR-UTE knee data, thereby accelerating MR-based 3D imaging of bones.

**Methods:** Data were acquired with a 3D radial IR-UTE pulse sequence, implemented using the open-source framework Pulseq. A denoising convolutional neural network (DnCNN) was trained in a supervised fashion using data from eight healthy subjects. Conjugate gradient sensitivity encoding (CG-SENSE) reconstructions of different retrospectively undersampled subsets (corresponding to 2.5-min, 5-min and 10-min acquisition times) were paired with the respective reference dataset reconstruction (30-min acquisition time). The DnCNN was then integrated into a Landweber-based reconstruction algorithm, enabling physics-based iterative reconstruction. Quantitative evaluations of the approach were performed using one prospectively accelerated scan as well as retrospectively undersampled datasets from four additional healthy subjects, by assessing the structural similarity index measure (SSIM), the peak signal-to-noise ratio (PSNR), the normalized root mean squared error (NRMSE), and the perceptual sharpness index (PSI).

**Results**: Both the reconstructions of prospective and retrospective acquisitions showed good agreement with the reference dataset, indicating high image quality, particularly for an acquisition time of 5 min. The proposed method effectively preserves contrast and structural details while suppressing noise, albeit with a slight reduction in sharpness.

**Conclusion**: The proposed method is poised to enable MR-based bone assessment in the knee within clinically feasible scan times.

# 1 Introduction

Over the past years, the visualization of bony tissue using magnetic resonance imaging (MRI) has been increasingly investigated. Both morphological imaging [1] and the quantification [2] of bony tissue have been explored. Typically, techniques such as radiography, computed tomography (CT) or scintigraphy are employed for the clinical imaging and assessment of bone. While these rely on ionizing radiation, MRI offers a radiation-free alternative. However, capturing the signal from bone is technically challenging due to its rapid decay, mainly characterized by the effective transverse relaxation time (T2*) of collagen-bound water protons [3], which is approximately 400 μs at 3T [4,5].

To acquire signals from tissues with such short T2, the echo time (TE) must be correspondingly short, leading to the development of ultrashort TE (UTE) imaging [6]. However, the proton density of bone is also low compared to surrounding tissues such as muscle and fat [4,7]. Fat, in particular, is a predominant component in trabecular bone [8]. Consequently, osseous tissue remains hypo-intense in standard UTE images. To address this, selectively suppressing long T2 signal components has proven beneficial in highlighting tissues of interest [9,10].

One technical approach is UTE imaging with echo subtraction. In this method, multiple echoes with different TE values are acquired and subtracted from one another, effectively suppressing long T2 signal components and enhancing the contrast of short T2 tissues [6,7,11]. This technique has been used, for example, to visualize cortical bone and ligaments in the knee [12]. The DURANDE sequence, an advanced UTE echo subtraction method utilizing two different radiofrequency (RF) excitation pulses to exploit the RF sensitivity of short T2 signals [13], has been applied for cranial bone imaging [14–16]. However, limitations of echo subtraction include sensitivity to B0 field inhomogeneities and the presence of multiple fat resonance peaks, which can lead to residuals from unwanted tissue after subtraction [9].

In adiabatic inversion recovery UTE (IR-UTE), a preparation pulse is applied, saturating the magnetization of tissues with T2 << d and fully inverting tissues with T2 >> d, where d represents the pulse duration of the adiabatic inversion pulse [9,17]. Signal acquisition begins at the inversion time (TI), where tissues with long T2 reach the zero crossing and do not contribute to the captured signal. Meanwhile, tissues with short T2 have significantly recovered, allowing for their selective visualization [18]. A key advantage of the IR-UTE sequence is the robustness of the adiabatic pulse against B0 and B1 field inhomogeneities [9,19].

IR-UTE sequences enable hyperintense imaging of both trabecular [8] and cortical [20–22] bony tissue across various anatomical regions. Additionally, quantitative metrics derived from IR-UTE images exhibit a significant correlation with the mechanical properties of bone [23] and quantitative values obtained from CT [24]. The latter, for example, offers a promising avenue for MR-based diagnosis of osteoporosis.

The primary drawback of this technique is the long acquisition time required to sufficiently encode an image. Only a limited number of lines in k-space can be acquired after applying the inversion pulse, and the magnetization must then recover before the next inversion. This limitation has hindered its adoption in clinical practice so far [8].

A pragmatic solution to reduce acquisition time is "undersampling", i.e., acquiring fewer projections than required by the Nyquist criterion (i.e., a "fully sampled" dataset). To maintain adequate image quality from undersampled data, acceleration techniques such as compressed sensing [25,26] or machine learning-based strategies [27–33] have been proposed.

This work presents a data-driven approach to accelerate 3D IR-UTE imaging of the knee. Specifically, an iterative reconstruction algorithm with an embedded denoising convolutional neural network (DnCNN) is introduced. The DnCNN was trained using prospectively acquired data collected from healthy volunteers specifically for this project.

## 2 Methods

### 2.1 IR-UTE Sequence

A 3D-IR-UTE pulse sequence (Fig. 1) was developed and implemented using the Pulseq [34] framework in MATLAB (The MathWorks, Natick, MA, USA). It is based on the 3D-IR-UTE-Cones sequence proposed in [8]. Initially, preparation is carried out using an adiabatic inversion pulse. This RF pulse follows a hyperbolic secant shape with a duration of 10 ms, a bandwidth of approximately 1.6 kHz, and a maximum nominal amplitude of 18.5 µT. It is centered around -220 Hz, which corresponds to the midpoint between fat and water at 3T. At TI = 64 ms, the inverted magnetization of long T2 components is expected to reach the zero crossing, while the saturated magnetization of short T2 tissues has significantly recovered [8,35].

To increase sampling efficiency, seven projections centered around TI were recorded per adiabatic IR pulse. A temporal spacing of τ = 3.8 ms between subsequent readouts and a repetition time of TR = 150 ms between successive inversions were used, following [8]. Data sampling was performed using short TE = 30 µs plus ramp sampling on a center-out 3D radial trajectory, which was evenly distributed over the k-space sphere (commonly known as koosh ball) using a Fibonacci lattice [36]. This sampling strategy enables rapid coverage of k-space periphery, maximizing signal-to-noise ratio (SNR) in high spatial frequency regions. The achieved resolution, based on the maximum k value, is approximately 1.3 mm isotropic. Non-selective rectangular excitation pulses with a duration of 40 µs, a nominal flip angle α = 18°, and RF-spoiling [37] were applied.

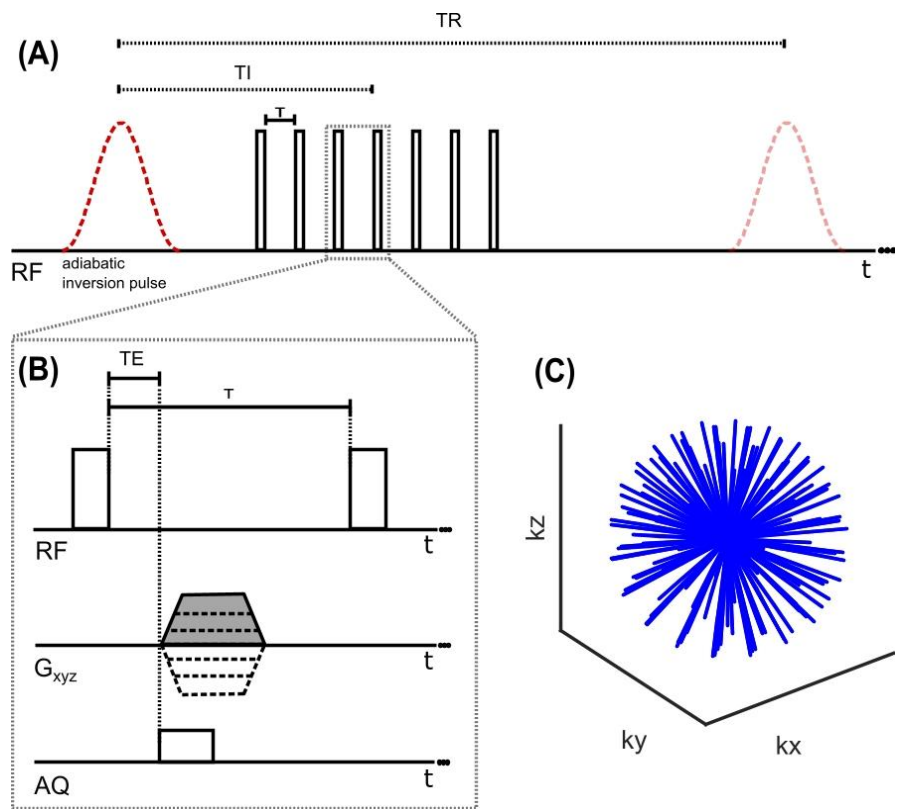

Figure 1: (A): Overview of 3D IR-UTE sequence. Adiabatic inversion pulse followed by seven projections with the central projection beginning at TI = 64 ms, where magnetization of long T2 is supposed to reach the zero crossing. The interval between two consecutive adiabatic IR pulses is TR = 150 ms. (B): Single excitation pulse and projection. τ = 3.8 ms denotes the time interval between two RF pulses. Ramp sampling and short TE= 30 µs are applied for optimal signal gain. (C): Schematic 3D representation of radial koosh ball trajectory.

## 2.2 In vivo bone study of the knee

The study was approved by the local ethics committee under license 122/23-sc, and written informed consent was obtained from each participant. Inclusion criteria required participants to be ≥ 18 years old and in healthy condition. Exclusion criteria included standard contraindications for MRI examinations.

A total of twelve healthy volunteers (pseudonymized as V1 to V12) were examined on a 3T clinical scanner (Magnetom Prisma$^{fit}$, Siemens Healthineers, Erlangen, Germany). Data were acquired in the left knee using a 15-channel transmit/receive knee coil. For all volunteers, a 30-min scan consisting of 84k projections was acquired, serving as the reference dataset in each case. For one volunteer (V12), additional scans with 7k projections (2 minutes 30 seconds) and 14k projections (5 minutes) were performed, following the same koosh ball trajectory and otherwise identical parameters. These acquisitions are referred to as "prospectively accelerated scans". Furthermore, UTE scans without IR preparation were conducted on the same subject using TR = 8 ms and α = 1°. These were performed with the same trajectory as their corresponding IR-UTE measurements for 7k (54 seconds) and 14k (1 minute 48 seconds) projections.

## 2.3 Training and test data

For all examinations (V1 to V12), undersampled subsets from the reference scan (84k projections) were generated by selecting every third (acceleration factor R = 3, 28k projections), sixth (R = 6, 14k), and twelfth projection (R = 12, 7k) (Fig. 2). These are referred to as "retrospectively undersampled datasets". All datasets were reconstructed using convolution gridding-based [38,39] conjugate gradient (CG)-SENSE [40] with a grid size of 256 x 256 x 256. Coil sensitivities were estimated individually from the densely sampled k-space center [41]. CG-SENSE reconstructions of eight of the examinations (V1 to V8) were exclusively used for training the neural network. The remaining four datasets (V9 to V12) were used for evaluation.

## 2.4 Network architecture and training

Due to the low proton density of the imaged tissue, IR-UTE images are characterized by a comparatively low SNR. This becomes particularly problematic for accelerated scans, as $SNR \sim \sqrt{T_{total}}$, where $T_{total}$ represents total sampling time. To address this, a dedicated denoising convolutional neural network (DnCNN) was trained to regularize an iterative MR image reconstruction of IR-UTE data.

The DnCNN architecture used is a modified version of [42] (Fig. 2) with an end-to-end skip connection. Following an initial single-channel 2D input convolutional layer (Conv) with 64 3 x 3 kernels and a rectified linear unit (ReLU) activation, six residual blocks were implemented [32,43,44]. Each residual block contains two sequences of Conv (64 3 x 3 kernel), batch normalization (BN), and ReLU with a skip connection. The final Conv, followed by ReLU and BN, is succeeded by a 1 x 1 Conv with a single filter to merge the features into a single channel. The total number of learnable parameters amounts to 482.9k.

To train the model separately for different acceleration factors, individual sagittal 2D slices from CG-SENSE reconstructions of the retrospectively undersampled datasets were paired with the corresponding reference images from the CG-SENSE reconstructions of the 30-min reference datasets. This resulted in 3,840 (R = 3), 7,680 (R = 6), and 15,360 (R = 12) input-target pairs. Data from volunteers V1 to V8 were converted to magnitude images, and normalized to a 0-1 range. The network was trained using only magnitude data, and the complex phase was estimated within the iterative reconstruction procedure (see below).

The network was implemented and trained in MATLAB (MathWorks, Natick, MA) utilizing the Deep Learning Toolbox. The training was conducted using 500 epochs with a batch size of 64, patch size of 80 x 80 and mean absolute error (MAE) as loss function. Adam [45] was used as the optimizer with an initial learning rate of $10^{-4}$ which was reduced by a factor of 0.5 every 100 epochs. The multi-GPU training utilized eight Nvidia Quadro RTX 6000 GPUs.

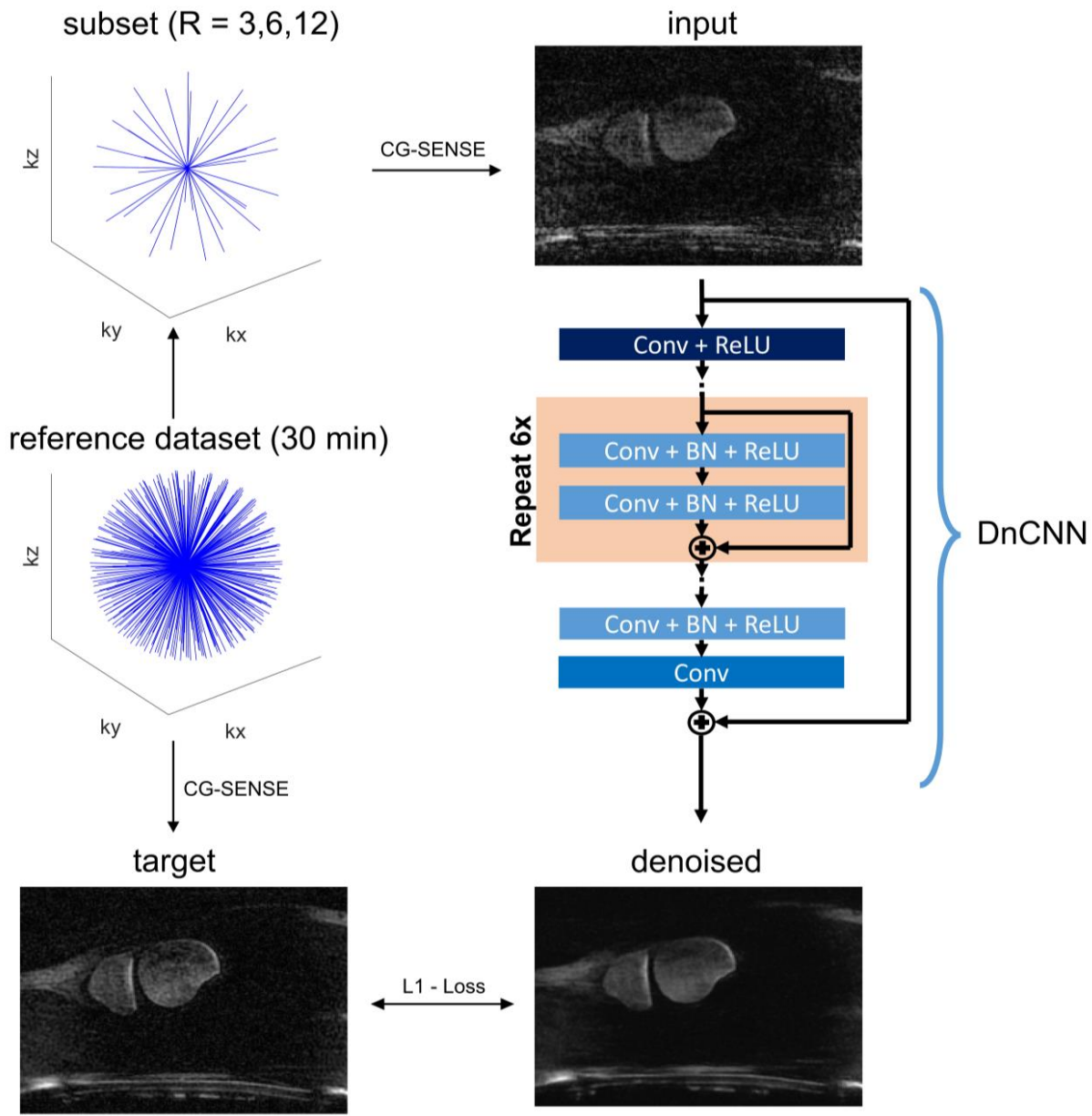

Figure 2: Schematic overview of training data creation and the network. From the acquired 30-min reference dataset, undersampled subsets were simulated by retrospectively removing projections (i.e., R=3, R=6 and R=12). CG-SENSE reconstructed subsets serve as inputs, paired with respective CG-SENSE reconstructions of the reference dataset as targets. The DnCNN features an end-to-end skip connection and six residual blocks each consisting of two series of convolutional layer (Conv), batch normalization (BN) and rectified linear unit (ReLU) with skip connection. Input and target slices were normalized and phase-shifted prior to training. Adam is used as the optimizer and MAE as the loss function.

## 2.5 Iterative Reconstruction algorithm

MRI reconstruction can be formulated as the inverse problem:

$$Ax = y,$$

where A denotes the MRI forward operator (including 3D gridding [38,39] and coil sensitivity estimation [41]), x represents the image data, and y the acquired k-space data. This ill-posed problem is addressed by minimizing the least squares error:

$$x = argmin_x \frac{1}{2} \|Ax - y\|.$$

A straightforward means to solve this problem is provided by the Landweber (LW) iterations [46]:

$$x_{m+1} = x_m - \xi A^*(Ax_m - y),$$

where $\xi$ is the step size and A* is the adjoint of A. To introduce a regularizer – in this case, the DnCNN – a plug-and-play approach [47] can be derived from the LW-iterations, enforcing both physical data consistency and a learned model. The following procedure was implemented for this purpose (please note, that the 2D DnCNN is applied sequentially on the individual sagittal slices here).

Pseudo-Code:

```
Input: x₀ = A*y, ξ = 1
for m = 0,1,2,…,n − 1
    Pₘ = angle(xₘ)
    z = max(abs(xₘ))
    xₘ = xₘ/z
    v = xₘ
    xₘ = real(xₘ ∘ e^(i·(−Pₘ)))
    xₘ = DnCNN(xₘ)
    xₘ = xₘ ∘ e^(i·Pₘ)
    xₘ = σ · xₘ + (1 − σ) · v
    xₘ = xₘ · z
    xₘ₊₁ = xₘ − ξ A*(Axₘ − y)
end
return xₙ
```

A total of n=6 iterations were used and the denoising weighting factor was set to $\sigma=0.15$. The initial input image $x_0$ was determined by "naïve gridding reconstruction" on a 256 x 256 x 256 grid.

The reconstruction algorithm was tested using retrospectively undersampled datasets from subjects V9 to V12 and the prospectively accelerated scans from volunteer V12. The proposed method is referred to as S3MOB (Speedy 3D MRI of Bone).

For comparison, reconstructions based on the retrospectively undersampled datasets were also performed using the "denoiseImage" function from MATLAB's Deep Learning Toolbox, replacing the custom DnCNN, while keeping all other parameters identical. This function applies a pre-trained DnCNN model based on [42], which is not specifically trained for IR-UTE-based MRI. This is referred to as LW-dIf (Landweber iterations with denoiseImage function).

### 2.6 Quantitative evaluation of image quality

To assess the image quality of the resulting reconstructions from the healthy volunteers' datasets, the following metrics were used (in each case, the reference image was used as the comparative value): structural similarity index measure (SSIM), peak signal-to-noise ratio (PSNR), normalized root mean squared error (NRMSE) as well as perceptual sharpness index (PSI) [48].

Metrics were obtained by normalizing and averaging within a 2D rectangular region-of-interest (ROI) across sagittal knee slices containing osseous tissue. The resulting volume-of-interest (VOI) included all bony structures of the respective knee. The final value for each metric was obtained by averaging the values derived from volunteers V9 to V12. Metrics were also calculated for CG-SENSE-Dn, where the custom DnCNN was applied directly to CG-SENSE reconstruction.

For the prospectively accelerated case (V12), images were also registered to the reference image prior to calculating the metrics using MATLAB and rigid transformation (translation and rotation).

## 3 Results

Training the different DnCNNs required approximately 10 hours (R = 3), 23 hours (R = 6), and 45 (R = 12) hours, respectively. The trained models were then used for reconstruction, with the reconstruction time for the proposed iterative method being approximately six minutes per 3D image.

Fig. 3 and Fig. 4 show sagittal and coronal slices of volunteer V9 obtained by applying the described reconstruction techniques for different levels of retrospective undersampling. Images reconstructed with the proposed S3MOB method exhibit a reduced noise level compared to CG-SENSE and LW-dIf reconstructions, particularly for higher undersampling. CG-SENSE-Dn achieves considerable noise suppression but introduces some blurring. Additionally, streaking artifacts can be observed in the coronal view, becoming more pronounced at higher acceleration levels.

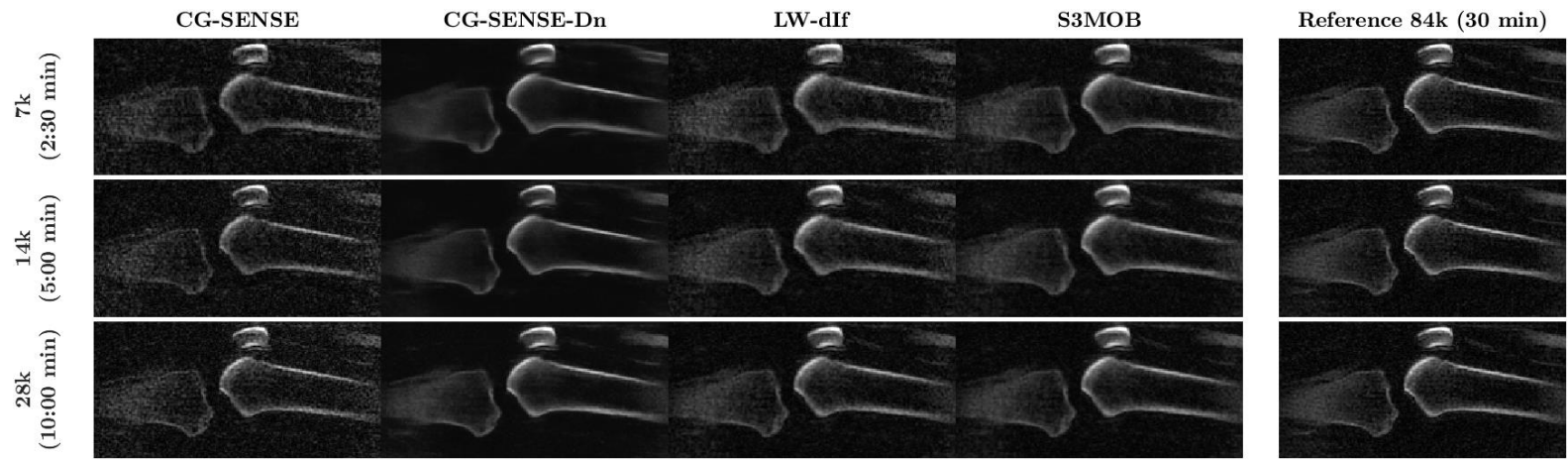

Figure 3: Sagittal knee slice of volunteer V9 to compare CG-SENSE reconstruction (CG-SENSE), single application of the respective custom DnCNN on the CG-SENSE reconstruction (CG-SENSE-Dn), iterative Landweber-based reconstruction including "denoiseImage" function (LW-dIf), and S3MOB for different number of retrospectively undersampled 3D IR-UTE datasets (corresponding acquisition time). The reference image obtained from the CG-SENSE reconstruction of the 30-minute reference dataset is shown in the last column. The proposed technique yields good suppression of noise while preserving contrast and detail.

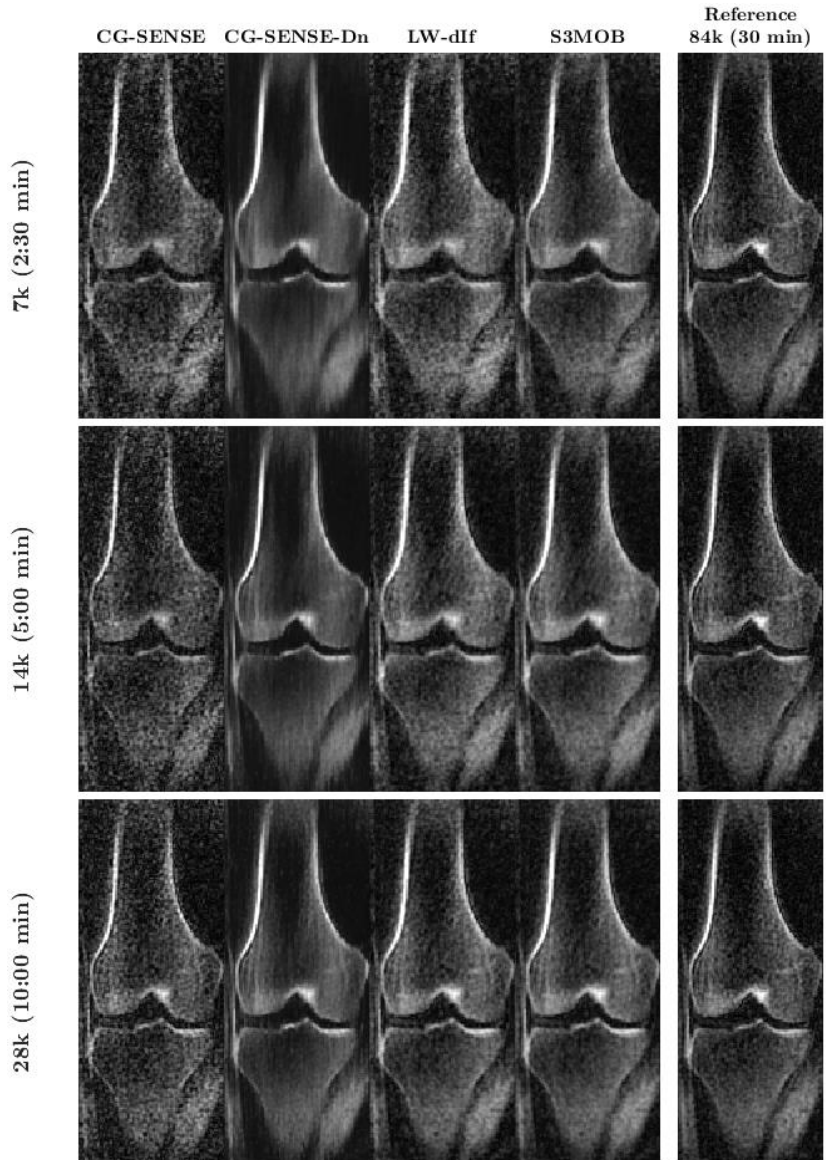

Figure 4: Coronal knee slice of volunteer V9 to compare CG-SENSE reconstruction (CG-SENSE), single application of the respective custom DnCNN on the CG-SENSE reconstruction (CG-SENSE-Dn), iterative Landweber-based reconstruction including "denoiseImage" function (LW-dIf), and S3MOB for different number of retrospectively undersampled 3D IR-UTE datasets (corresponding acquisition time). The reference image obtained from the CG-SENSE reconstruction of the 30-minute reference dataset is shown in the last column. The proposed technique yields good suppression of noise while preserving contrast and detail. For CG-SENSE-Dn, striation artifacts due to blurring can be observed, which could be explained by the training of the DnCNN on sagittal slices. S3MOB is capable of minimizing the occurring striation artifacts along with the blurring.

Table 1 presents the image quality metrics obtained from volunteers V9 to V12. These values are reported for different reconstruction methods and levels of retrospective undersampling. Across all reconstruction methods, the metrics tend to improve with increasing amount of data, except for PSI in the iterative reconstruction methods, where the values remain comparable across different undersampling levels. CG-SENSE reconstructions result in the highest PSI for each level of undersampling, indicating better sharpness compared to other reconstruction methods. However, this comes at the cost of the poorest similarity metrics (i.e., lower SSIM, lower PSNR, and higher NRMSE). Conversely, CG-SENSE-Dn achieves the best similarity values compared to other methods but at the expense of sharpness, as indicated by the lowest PSI.

| Method | Time | SSIM | PSNR [dB] | PSI | NRMSE |
|---|---|---|---|---|---|
| CG-SENSE | 2.5 min | 0.767 (0.029) | 33.82 (0.65) | 0.400 (0.020) | 0.414 (0.028) |
| | 5 min | 0.770 (0.025) | 33.93 (0.46) | 0.432 (0.019) | 0.409 (0.030) |
| | 10 min | 0.771 (0.028) | 34.07 (0.65) | 0.441 (0.009) | 0.405 (0.067) |
| CG-SENSE-Dn | 2.5 min | 0.876 (0.020) | 35.83 (1.03) | 0.218 (0.030) | 0.330 (0.043) |
| | 5 min | 0.892 (0.016) | 36.90 (0.69) | 0.264 (0.015) | 0.291 (0.033) |
| | 10 min | 0.907 (0.014) | 38.38 (0.39) | 0.288 (0.031) | 0.245 (0.024) |
| LW-dIf | 2.5 min | 0.800 (0.027) | 34.03 (0.83) | 0.361 (0.027) | 0.408 (0.074) |
| | 5 min | 0.839 (0.027) | 34.93 (1.23) | 0.344 (0.021) | 0.371 (0.087) |
| | 10 min | 0.872 (0.028) | 36.01 (1.64) | 0.356 (0.016) | 0.331 (0.093) |
| S3MOB | 2.5 min | 0.834 (0.025) | 35.00 (0.97) | 0.323 (0.024) | 0.366 (0.073) |
| | 5 min | 0.858 (0.022) | 35.29 (1.23) | 0.306 (0.024) | 0.356 (0.084) |
| | 10 min | 0.890 (0.019) | 36.61 (1.48) | 0.315 (0.028) | 0.307 (0.080) |
| Reference | 30 min | 1 | ∞ | 0.394 (0.038) | 0 |

Table 1: Image quality metrics as average (standard deviation) obtained from the retrospectively undersampled data of volunteers V9 to V12 (who were not included in the DnCNN training data). The time indicates the corresponding acquisition time.

In comparison to CG-SENSE and CG-SENSE-Dn, the iterative methods yield intermediate values. The proposed S3MOB method demonstrates higher similarity values (except for NRMSE in the 10-min undersampled dataset) compared to LW-dIf. In contrast, LW-dIf achieves higher PSI values.

Fig. 5 and Fig. 6 illustrate a comparison between reconstructions of prospectively and retrospectively undersampled datasets obtained from V12 using the proposed method for 2.5-min and 5-min acquisition times. Note that the unregistered prospectively undersampled images do not show exactly the same region as the reference images, as small deviations due to movement between scans cannot be excluded. Overall, all images successfully visualize morphological structures of the bones in the knee. Retrospective images tend to exhibit slightly better image quality, confirmed by higher metric values as shown in Table 2.

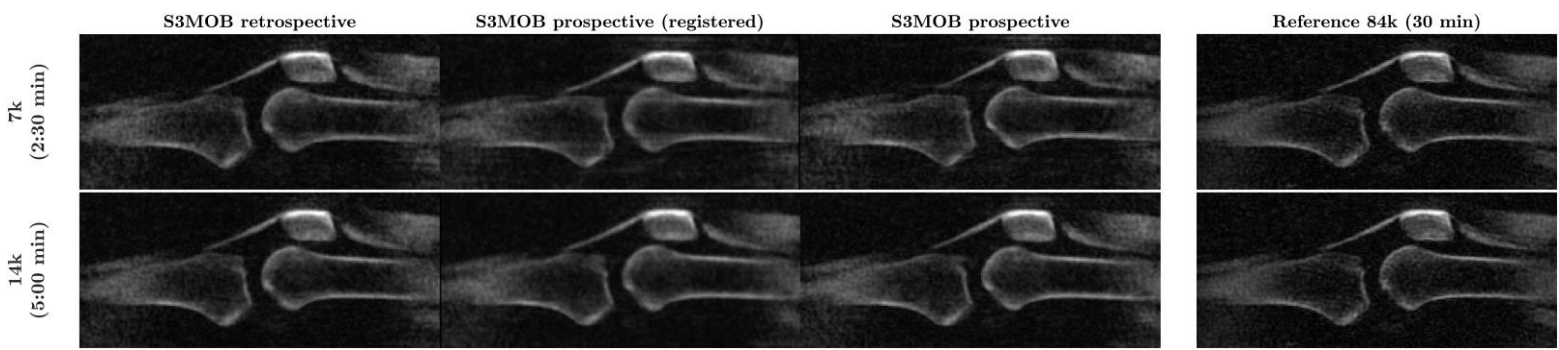

Figure 5: Sagittal knee slice of volunteer V12 reconstructed from retrospectively undersampled datasets and prospectively accelerated scans with the proposed method (prospective, with and without registration). Vertically, two different acceleration factors or acquisition durations are given. Note that the unregistered prospective slices do not show exactly the same region, most likely due to movement of the volunteer during the scan. The registered prospective images were aligned to the reference image after reconstruction. The images clearly depict the bony structure of the knee joint, including the kneecap. In the 5-min prospective images, no streaking artifacts are visible compared to those obtained from 2.5-min scan time.

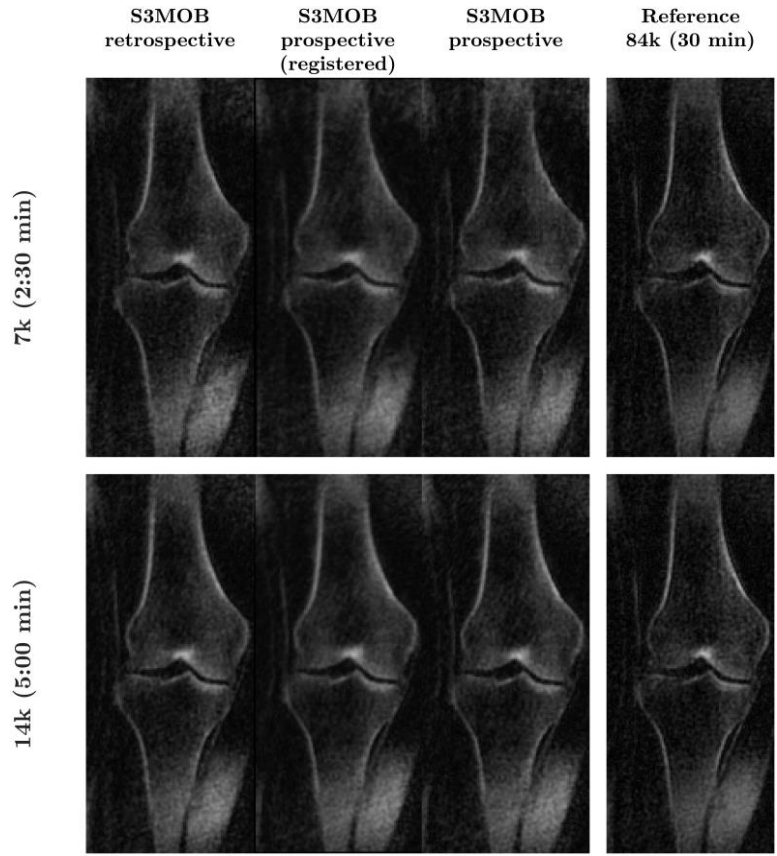

Figure 6: Coronal knee slice of volunteer V12 reconstructed from retrospectively undersampled datasets and prospectively accelerated scans with the proposed method (prospective, with and without registration). Vertically, two different acceleration factors or acquisition durations are given. Note that the unregistered prospective slices do not show exactly the same region, most likely due to movement of the volunteer during the scan. The registered prospective images was aligned to the reference image after reconstruction. The images distinctly visualize the joint space between the femur and tibia, particularly in the prospective case with 5-min acquisition time.

Registration of prospectively undersampled images improves similarity values for the respective acquisition time but comes at the cost of lower PSI values. This effect is particularly noticeable at the higher acceleration (2.5-min acquisition time), where the images exhibit slight blurring compared to the 5-min case and retrospectively undersampled images. Additionally, in the sagittal slice, moderate streaking artifacts can be observed. The 5-min prospectively undersampled sagittal image shows a structure on the tibia that is not visible in the other images. This structure may be part of the anterior crucial ligament (ACL), which is not yet visible in other slices due to the mentioned orientation deviations.

| S3MOB | Time | SSIM | PSNR [dB] | PSI | NRMSE |
|---|---|---|---|---|---|
| prospective | 2.5 min | 0.739 | 31.89 | 0.300 | 0.494 |
| prospective (registered) | 2.5 min | 0.794 | 33.88 | 0.254 | 0.393 |
| prospective | 5 min | 0.807 | 33.92 | 0.313 | 0.391 |
| prospective (registered) | 5 min | 0.837 | 35.22 | 0.233 | 0.337 |
| retrospective | 2.5 min | 0.826 | 34.73 | 0.309 | 0.356 |
| retrospective | 5 min | 0.857 | 35.78 | 0.319 | 0.316 |
| Reference | 30 min | 1 | $\infty$ | 0.415 | 0 |

Table 2: Image quality metrics (standard deviation) obtained from reconstructions of the prospectively accelerated and retrospectively undersampled data for volunteer V12 using the proposed S3MOB method (prospective, with and without registration). In case of prospective acceleration, the time indicates the actual acquisition time. In case of retrospective undersampling, the time indicates the corresponding acquisition time.

# 4   Discussion

In this work, we explored the feasibility of accelerating radial IR-prepared 3D UTE by reconstructing undersampled acquisitions using a physics-based iterative reconstruction with an embedded DnCNN specifically trained for this purpose.

Regarding the metrics and qualitative image impression, the application of our proposed S3MOB technique yields improved similarity scores (SSIM, PSNR, NRMSE) with respect to the reference for retrospectively and prospectively accelerated acquisitions, compared to CG-SENSE and LW-dIf. S3MOB also exhibits significantly higher PSI values, indicating better sharpness compared to CG-SENSE-Dn. However, this comes at the cost of lower sharpness (PSI) compared to CG-SENSE and LW-dIf, as well as poorer similarity compared to CG-SENSE-Dn. The latter can be explained by the fact that the DnCNN was optimized to achieve high similarity between the reconstruction of the respective undersampled dataset and its reference. This comes at the expense of introducing blurring, which is typical for denoising applications [32], most likely because generic loss-functions do not heavily penalize slight degradations in spatial resolution.

To mitigate blurring and preserve sharpness, the use of the tailored DnCNN within the reconstruction algorithm was weighted by $\sigma = 0.15$. A higher value of $\sigma$ is expected to increase nominal similarity but would likely reduce sharpness. Throughout the iteration process, it was observed that the data consistency tends to preserve sharpness.

In general, the proposed method successfully balances resolution and denoising, where fine-tuning the weighting and the number of iterations could further enhance performance. These two parameters were held constant across all S3MOB reconstructions, indicating its robustness. However, tuning these parameters individually for each volunteer could further optimize performance in specific cases.

A radial center-out sampling scheme was chosen to efficiently capture short bone signals, optimize contrast, and reduce blurring, at the cost of suboptimal k-space filling compared to spiral acquisition [49]. The CG-SENSE reconstructions include data from an acquisition window time of 300 μs for undersampled datasets and 320 μs for reference datasets, which aligns with the optimal SNR range for radial UTE sampling [11].

This study has several limitations. The DnCNN was trained on data that includes not only bony structures of the knee, but also incorporates peripheral elements, such as coil components and the patient table. Additionally, unsuppressed parts of the leg are included due to the B1 sensitivity of the coil. This is evident when comparing UTE reconstructions with the corresponding IR-UTE reconstructions (see Supplementary Fig. 1). Furthermore, in prospective imaging, parts of the patient table appear brighter than in the retrospective images, especially at higher undersampling levels.

Another discrepancy between retrospective and prospective acquisitions could arise from signal contributions of different T2 compartments. Multiple spokes are acquired around the point where the z-magnetization of long T2 tissues is expected to reach the null point for the given TR/TI combination [8]. In the prospective acquisition, seven projections per IR-pulse are acquired consistently per TR. Given the small flip angle ($\alpha = 18$ °), excitation pulses have negligible impact on the longitudinal relaxation process of long T2 tissues (e.g., fat) [35]. This enables signal suppression for multiple projections per TR as signals acquired from spokes before the null point are of opposite sign compared to those acquired afterward, likely leading to cancellation during image reconstruction [9,35]. However, this is not the case for retrospectively undersampled datasets, where only every third, sixth, or twelfth projection is used from the combined trajectory. Nonetheless, since the total number of spokes is large (84k), mutual averaging can be assumed. To further enhance the equivalence between prospectively and retrospectively undersampled acquisitions, acquiring only one or fewer spokes per inversion pulse would be beneficial. However, this would, in turn, increase scan time for the same TR and total number of spokes.

Additionally, patient movement during scans may have impaired the evaluation, particularly when comparing prospectively accelerated with retrospectively undersampled reconstructions or with their respective references. For volunteer V12 who underwent prospectively undersampled scanning, the total examination duration was approximately 1 hour, during which involuntary movement cannot be entirely ruled out. This might explain why the registration of prospective images improves similarity metrics compared to unregistered images, while simultaneously decreasing PSI, indicating that blurring was introduced by the registration optimizer.

Furthermore, the training data originated from only eight healthy volunteers. Improved DnCNN performance and, consequently, better reconstruction quality, is expected with more training data, particularly by including patient datasets, which could enhance pathology detection in later applications. Notably, the first four volunteers (V1 to V4) used for training and one volunteer used for evaluation (V10) were accidentally acquired with a suboptimal trajectory in terms of k-space uniformity. However, since the primary task of the method was denoising rather than removing undersampling artifacts, and because no qualitative differences were observed in naïve reconstructions compared to images obtained with the optimized trajectory, these datasets were still included in the training and evaluation, respectively.

Future studies could investigate patients with tibial plateau fractures to test a one-stop-shop approach for fracture morphology and soft tissue assessment in MRI. If successful, additional CT imaging may become absolute in these individuals. Additionally, this technique could be extended to other anatomies, such as the lumbar spine, where the effect of accelerated scanning on collagen-bound water proton fraction (CBWPF) maps is of particular interest. CBWPF has been shown to significantly correlate with CT-based bone mineral density (BMD) and represents a potential biomarker for MR-based osteoporosis assessment [24].

Moreover, training the model using a variational network approach [50] could optimize the network at each iteration, potentially leading to an improved trade-off between noise and resolution. Furthermore, transitioning to a fully 3D approach would be highly desirable and could be a focus of future work.

# 5 Conclusion

We presented an iterative Landweber-based reconstruction algorithm incorporating a plug-and-play DnCNN to reconstruct undersampled radial 3D IR-UTE datasets of the knee. The DnCNN was trained using CG-SENSE reconstructed images obtained from healthy volunteers across different levels of undersampling. Good image quality and high similarity to the reference datasets were achieved while preserving sharpness, contrast, and detail for both prospective acceleration and retrospective undersampling. This technique demonstrates potential for MR-based bone assessment within a clinically feasible scan time.

## Declaration of competing interest

The authors declare that they have no known competing financial interests or personal relationships that could have appeared to influence the work reported in this paper.

## Data availability

Participants were assured that the raw data would be used exclusively for the purposes of this study.

## Supplementary Material

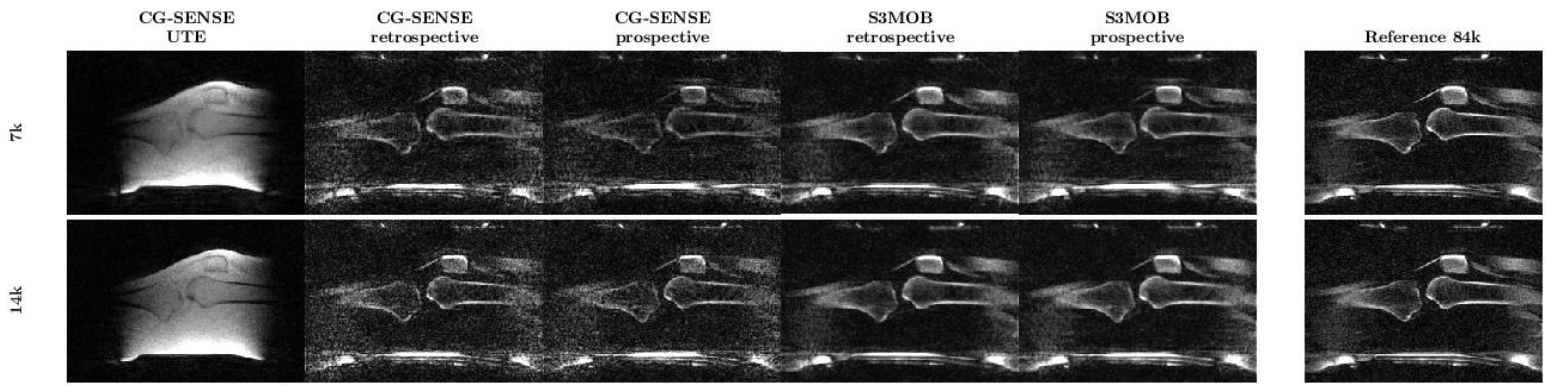

Supplementary Figure 1: Sagittal knee slice of a healthy volunteer reconstructed from UTE (first column) and IR-UTE (other columns) with different methods. 7k and 14k spokes corresponds to 54 seconds and 1 minute 48 seconds acquisition time for UTE and 2 minutes 30 seconds and 5 minutes for IR-UTE acquisitions, respectively. When comparing the reconstructed UTE image with one of the IR-UTE images, unsuppressed regions of the legs are notably visible in the UTE image, particularly in areas where the coil appears less sensitive, as indicated by the UTE images. Additionally, in the prospectively acquired IR-UTE images, the patient table and parts of the coil exhibit higher signal intensities. This might influence the normalization step in the algorithm and adversely affect the reconstruction.